# Blockchain-based Payment Systems: A Bibliometric & Network Analysis

Shlok Dubey

**Abstract**
Blockchain is a shared, immutable ledger that has attracted the attention of researchers and practitioners across innumerable sectors, with its implications for modernizing payment systems having the possibility of inciting a digital revolution. In the scope of this study, 1,511 publications were obtained from Scopus to conduct a systematic review of the research space through bibliometric and network analyses. The main aim of this study was to determine key authors, significant studies, and collaboration patterns, to reveal the distributions and impacts of publications in the blockchain-based payments area between 2019 and 2022. The results indicate that the Khalifa University of Science and Technology is the most influential journal, while the most cited author is Salah, K. Additionally, the National Natural Science Foundation of China has sponsored most academic documents, with the US emerging as the most impactful country. This study has also found that blockchain-based payments literature congregated in 5 disciplines. These areas are computer science, engineering, decision science, mathematics, and business. A co-authorship analysis also networks relations between nations, authors, and organizations globally, creating unique clusters that maximize research productivity. In summary, this study designs an analytical map of the research landscape, which can guide future research.

## 1. Introduction

To appreciate the impact of blockchain-based payment systems, it is important to understand how blockchain operates. In essence, blockchain is a distributed and immutable digital ledger that stores information across a network (Hayes 2022; Rodeck 2022). Blockchain emerged when Stuart Haber and W. Scott Stornetta wrote about the implications of a cryptographically secure chain of blocks (Iredale 2020). Then, in 2008, Satoshi Nakamoto built on this concept in his whitepaper: 'Bitcoin: A Peer-to-Peer Electronic Cash System' (Sarkar 2021). He applied blockchain to payment, conceptualizing cryptocurrency for the world (Hayes 2022).

Blockchain is made up of a growing chain of blocks. Each block is a digital ledger that builds on the block before it. Each block also contains a hash that acts as its unique identifier. Blocks are cryptographically linked to one another as block 2's previous hash is block 1's hash. This ensures immutability as discrepancies between these two numbers indicate the possibility of a hack. Blockchain runs on the SHA-256 algorithm (Secure Hash Algorithm). This algorithm is a patented cryptographic hash function that outputs a 256 bits long value (N-able 2019). This hexadecimal hash is what later becomes a block's fingerprint.

A blockchain network must possess a few key characteristics to be considered operable. First is the avalanche effect. In cryptography, the avalanche effect means that a minimally changed input should produce a completely different output (Cryptovision 2021). Furthermore, blocks must be one-way–-using a hash to reverse engineer the information stored should be impossible; blockchain is a way to identify data, not retrieve it. Thirdly, it must be deterministic. This means that the initial condition should always produce the same results. Deterministic systems allow the network to have a consensus.

Blockchain has achieved the unthinkable. It has created trust in a trustless network. Thus, it is no surprise that it can be found in countless disciplines with numerous applications. This paper speaks on its applications in payment.

Some advantages of blockchain in payments are that it offers transparency, safe and quick payments, automation with smart contracts, and no intermediaries. Removing intermediaries allows transactions to be settled quicker, with fewer chargeable services (Marino 2016). Furthermore, blockchain's immutability and public network access increase transparency. Additionally, cross-border payments like remittances are quicker, as blockchain reduces payment processing times (Sun et al. 2020). Lastly, Smart contracts can allow for instant payments and the automation of payment flows ("Using Smart Contracts for Automated Financial Features, Like Debt Repayment • Sila").

Use cases of payment systems in blockchain can be seen through trade finance, digital identity verification, P2P transfers, and cross-border payments (Klebanov 2021; Elsaid 2020). Blockchain makes digital identity verification more secure as it is immutable, and the authenticity of data is ensured. In trade finance, there are no chances for manual errors.

With blockchain's potential to incite a digital revolution, researchers have authored many bibliometric analyses to map the field; notably, (Gao et al. 316-332 2021), (Zeng et al. 102-107 2018), (Darabseh and Martins 2020), (Kamran et al. 2020), (Duan and Guo 2021) and (Nasir et al. 989-1004, 2021), have authored impactful publications about blockchain. Some scholarly papers, like (Darabseh and Martins. 2020), even address blockchain's applications to a specific discipline.

The blockchain industry is catching global attention, and its implications on modernizing payment systems are more critical now than ever. This is why academic papers like (Papadis and Tassiulas 227596-227609 2020) and (Tasatanattakool and Techapanupreeda 473-475 2018) articulate challenges and applications in scaling adoption. However, currently, there is neither a comprehensive review of the relationships between research constituents nor an idea of the trends. This bibliometric analysis identifies gaps in the space and points researchers in the right direction moving forward.

## 2. Research Methodology

A bibliometric analysis is used to summarize large quantities of data to understand the intellectual landscape of a given field ("Subject and Course Guides: Bibliometric Analysis and Visualization: Bibliometrics," 2022). In this paper, the analysis will yield emerging trends in academic documents on blockchain-based payment systems. This quantitative analysis will evaluate 1,511 educational documents and provide interpretations of their impacts on the space. Furthermore, through thorough performance analysis–the contributions of research constituents– and science mapping –the relationships between research constituents–this study aims to identify knowledge gaps and derive novel ideas for investigation (Donthu et al. 2021).

This paper builds on a research methodology template coined by Tchangalova and Coalter; they endorsed a four-step method to identify influential and future studies in a subject area. The four steps are as follows: "identify the research questions, define the boundaries, search and select the studies and bibliometric analysis, and present the results" (Tchangalova and Coalter 2020, cited by Moosavi et al. 2021). However, this paper also conducts a co-authorship analysis. Figure 1 demonstrates these modifications.

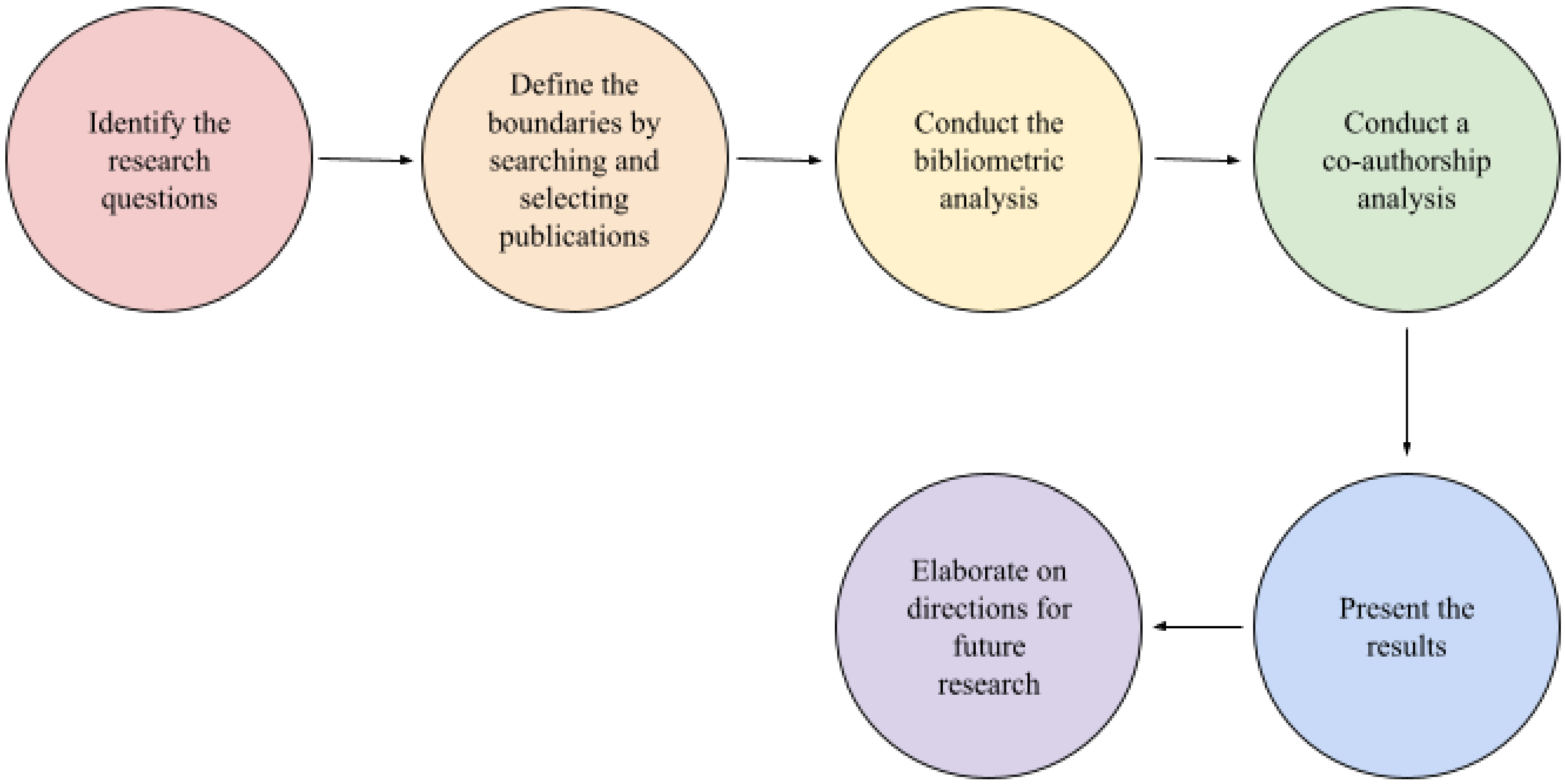

**Figure 1.** Research methodology. Google Drawings was used to create this graphical representation.

*2.1 Research Questions and Boundaries*

Previously, bibliometric analyses have been conducted to evaluate the blockchain research trend or look at blockchain applications in management (Tandon et al. 1, 2021; Miau and Yang 1 2018). Similarly, researchers, who have worked in the Cogent Business and Management Journal, authored literature reviews regarding mobile payment adoption (Naved Khan and Kostadinova 1 2021). However, this study is distinctive as it looks specifically at blockchain-based payments, providing a comprehensive overview of the research area.

Moreover, to demonstrate that this study adds new information to existing literature, the findings should be consistent with the following three research questions. (Alvesson and Sandberg 49, 2013).

- RQ1. How has literature on blockchain-based payment systems developed between 2019 and 2022?
    - RQ1.1. What are the most influential studies, affiliations, and authors?
    - RQ1.2. What are the geographical distributions and effects of publications over time?
    - RQ1.3. Which funding sponsors have the most extensive contributions?
- RQ2. What are the most important topics in blockchain-based payments literature?
- RQ3. What are the relationships among authors, countries, and organizations in blockchain-based payments literature?

Scopus databases will provide statistical information for authors, geographical distributions, affiliations, and funding sponsors. For RQ2, Scopus databases will provide a general idea of the distributions of topics. Moreover, a keyword co-occurrence analysis conducted through VOSviewer will recognize the most influential subjects. After cross-referencing this information and deducing key relationships, important topics in the research field will reveal themselves. Lastly, for RQ3, relationships among authors, countries, and organizations will be found through the VOSviewer clustering algorithm.

*2.2 Search and Selection*

About 99.11% of the journals indexed on the Web of Science are also indexed in Scopus (Singh et al. 2021). Thus, it is clear that Scopus encompasses enough scholarly papers to produce a thorough analysis. An article title, abstract, and keyword search of the words “blockchain” AND “payments” resulted in 1,870 document results. However, to maintain relevance, another constraint was placed, which excluded papers before 2018 and the 7 papers in 2023. This lowered the number of documents by 16.33% to 1,575 documents. A final constraint excluded Articles in Press, of which there were 64. Only publications in the final stage would be selected, leaving 1,511 papers. To maintain a diverse portfolio, non-English papers were included as well.

## 4. Bibliometric Analysis

This section presents the study’s findings based on the information gathered from the applied methodology. This section will not provide any interpretations of the data and will be arranged in a logical sequence corresponding to the research questions above (Sacred Heart University 2022).

For RQ1, there will be an analysis of the publications by year, university affiliations, geographical distributions, funding sponsors, most contributing and most cited authors, and the document type. For RQ2, a study will be conducted regarding the distribution of disciplines, author and index keyword co-occurrence, and most cited documents.

*4.1 Publication Trend by Year*

Of the 1,511 document types surveyed, 370 were published in 2019, 429 were published in 2020, 456 were published in 2021, and 256 were published in 2022. Figure 2 graphically represents this data.

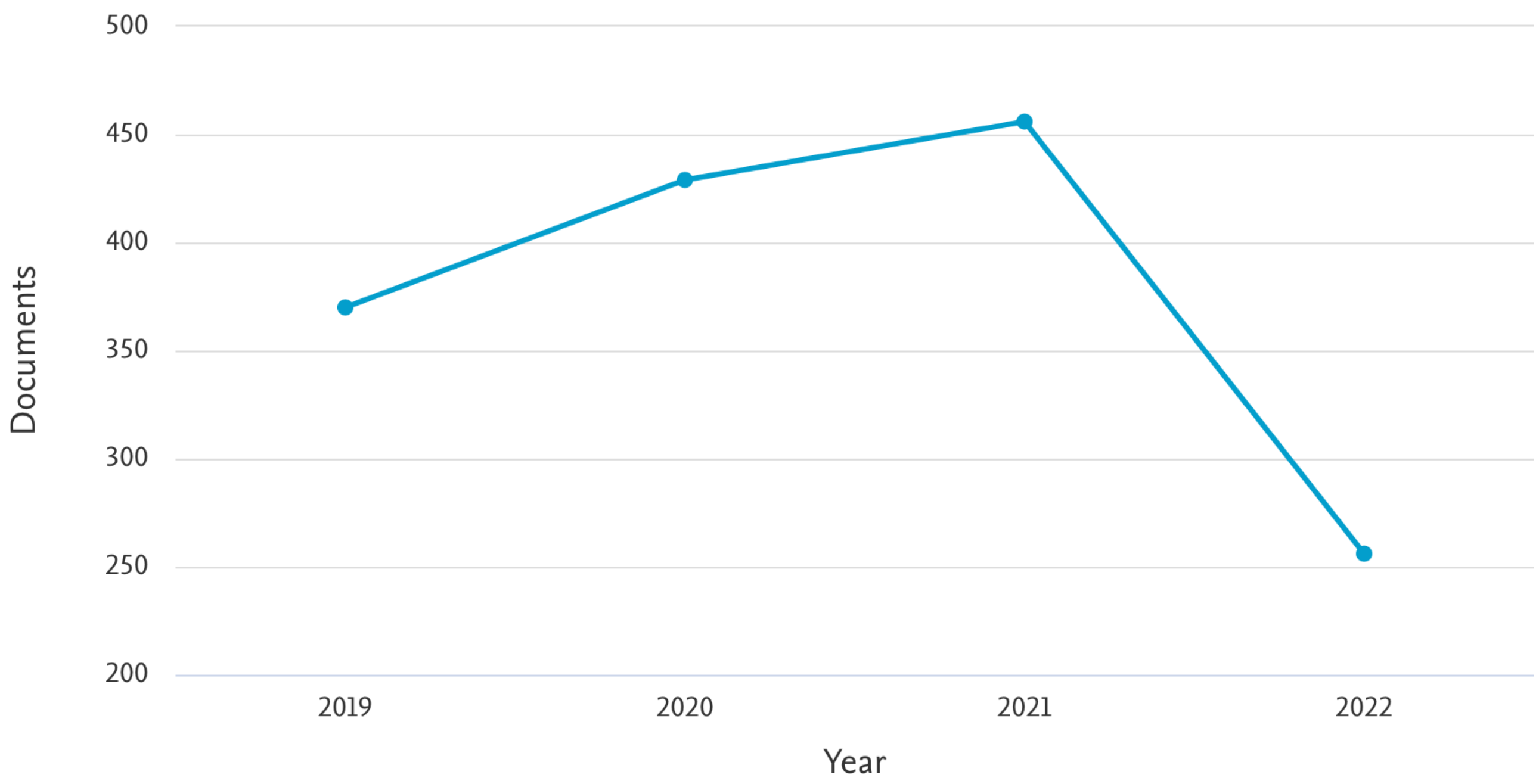

**Figure 2.** This is the publishing trend for educational documents (in the final stage) concerning blockchain-based payments between 2019 and 2022. This data and the graph were retrieved through the Scopus database.

While there was a 15.95% increase in publications between 2019-2020 and a 6.29% increase between 2020-2021, there was a 43.86% decrease in publications from 2021-2022 (as of October 2nd).

*4.2 Documents by Affiliation*

Table 1 shows the top ten affiliations with the most contributions to blockchain-based payment research. Table 1 also shows the TP (Total Publications), TC (Total Citations), and CPP (Citations per Publication). Scopus Databases provided the data for the TP and CP. TC/TP = CPP is the formula used to calculate CPP ("What are 'publication citations'?: Dimensions," 2021). The CPP measures the impact, per paper, on average, of an organization (Li and Ho 2007). The CPP is a more accurate measurement of an impact than the TC because it does not allow the TP to skew the data. Lastly, the h-Index, as defined by (Hirsch 2005), is "the number of papers with citation number ≥h, as a useful index to characterize the scientific output of a researcher." It measures the citation impact and productivity of a set of publications ("BeckerGuides: Tools for Authors: What is the h index?" 2022).

**Table 1.** Top 10 affiliations contributing to blockchain-based payment literature

| No. | Affiliation | TP | TC | CPP | h-Index |
|---|---|---|---|---|---|
| 1 | Chinese Academy of Sciences | 24 | 161 | 6.71 | 7 |
| 2 | ETH Zürich | 18 | 113 | 6.28 | 6 |
| 3 | Khalifa University of Science and Technology | 16 | 664 | 41.50 | 10 |
| 4 | Florida International University | 14 | 100 | 7.14 | 6 |
| 5 | Beijing University of Posts and Telecommunications | 14 | 37 | 2.64 | 3 |
| 6 | CSIRO Data61 | 14 | 70 | 5.00 | 4 |
| 7 | University of Electronic Science and Technology of China | 13 | 22 | 1.69 | 3 |
| 8 | SRM Institute of Science and Technology | 13 | 37 | 2.85 | 3 |
| 9 | Monash University | 13 | 61 | 4.69 | 4 |
| 10 | University of Chinese Academy of Sciences | 13 | 78 | 6.00 | 4 |

Note(s): This table was created with a dataset from Scopus

Six of these ten affiliations are located in Asian countries. The most impactful institutions measured by citations per publication are the Khalifa University of Science and Technology, Florida International University, and the Chinese Academy of Sciences. On the other hand, the most productive affiliations measured by the h-Index are the Khalifa University of Science and Technology, and the Chinese Academy of Sciences. The Khalifa University of Science and Technology, an affiliation highlighted throughout this paper, is based in the UAE and known for its 36 science, engineering, and medicine departments. They have been recognized as the most published in Blockchain Oracle research, giving them precedence for working with blockchain-based payments (Sterling 2021).

*4.3 Geographical Distribution*

Figure 3 is a graphical representation of the number of publications each country has authored concerning blockchain-based payments literature. This graph is a linear interpolation with a key that ranges from light green to dark blue. Progression across this gradient indicates an increase in academic papers published. Figure 3 includes data for 91 countries and 1,442 of the 1,511 documents studied (69 papers have undefined locations).

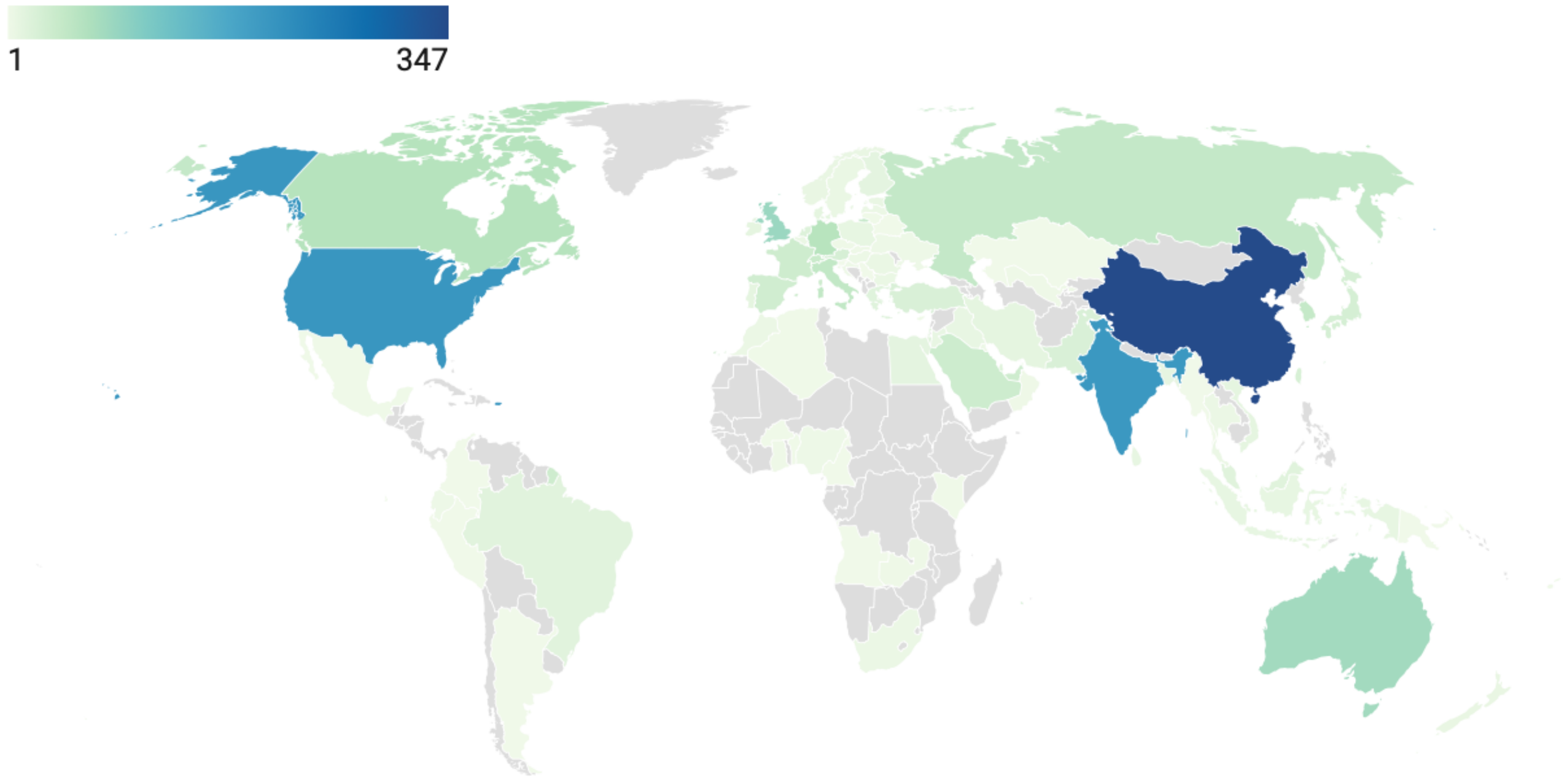

**Figure 3.** These are the geographical locations of contributing countries for educational documents (in the final stage) concerning blockchain-based payments between 2019 and 2022. This data was retrieved through the Scopus database, and the graph was created through Datawrapper.

Table 2 shows quantified statistics for the number of TP, TC, CPP, and the h-Index for the top 10 contributing countries.

**Table 2.** Top 10 countries contributing to blockchain-based payment literature

| No. | Country | TP | TC | CPP | h-Index |
|---|---|---|---|---|---|
| 1 | China | 347 | 2372 | 6.84 | 24 |
| 2 | United States | 208 | 2787 | 13.40 | 25 |
| 3 | India | 204 | 1583 | 7.76 | 16 |
| 4 | The United Kingdom | 84 | 960 | 11.43 | 14 |

| No. | Country | TP | TC | CPP | h-Index |
|---|---|---|---|---|---|
| 5 | Australia | 77 | 690 | 8.96 | 13 |
| 6 | Canada | 59 | 513 | 8.69 | 13 |
| 7 | Germany | 57 | 383 | 6.72 | 11 |
| 8 | Italy | 47 | 188 | 4.00 | 9 |
| 9 | Russian Federation | 45 | 283 | 6.29 | 7 |
| 10 | Republic of Korea | 44 | 283 | 6.43 | 9 |

Note(s): This table was created with a dataset from Scopus

The countries with the most average impact per document measured by CPP are the United States, the United Kingdom, and Australia. On the other hand, the most productive countries measured by the h-Index are the United States, China, and India.

*4.4 Documents by Funding Sponsors*

Table 3 shows the ten institutions that sponsored the most papers (in the final stage) about blockchain-based payments between 2019 and 2022.

**Table 3.** Top 10 funding sponsors contributing to blockchain-based payment literature

| No. | Funding Sponsor | TP | TC | CPP | h-Index |
|---|---|---|---|---|---|
| 1 | National Natural Science Foundation of China | 155 | 1550 | 10.00 | 21 |
| 2 | National Science Foundation | 44 | 442 | 10.05 | 11 |
| 3 | National Key Research and Development Program of China | 37 | 343 | 9.27 | 10 |
| 4 | Horizon 2020 Framework Programme | 26 | 226 | 8.69 | 9 |
| 5 | European Commission | 21 | 224 | 10.67 | 6 |
| 6 | National Research Foundation of Korea | 19 | 173 | 9.11 | 8 |
| 7 | Fundamental Research Funds for the Central Universities | 17 | 212 | 12.47 | 8 |
| 8 | European Research Council | 14 | 74 | 5.29 | 4 |
| 9 | Ministry of Science and Technology, Taiwan | 12 | 107 | 8.92 | 5 |
| 10 | National Basic Research Program of China (973 Program) | 11 | 106 | 9.64 | 5 |

Note(s): This table was created with a dataset from Scopus

The National Natural Science Foundation of China sponsors significantly more papers than other organizations as its related papers make up 10.26% of all blockchain-based payments literature. 5 out of 10 of the institutions in Table 3 are located in East Asian countries. The institutions with the most impactful publications, on average, as measured by CPP, are the Fundamental Research Funds for the Central Universities, the European Commission, and the National Science Foundation. On the other hand, the most productive sponsors measured by the h-Index are the National Natural Science Foundation of China, the National Science Foundation, and the National Key Research and Development Program of China.

### *4.5 Author Influence*

Table 4 shows the authors with the most publications (about blockchain-based payments). In this section, a new variable is measured: self-citations. A self-citation is when an author references their previously published work ("Self-citation and self-plagiarism, Research").

**Table 4.** Top 10 contributing authors to blockchain-based payment literature

| No. | Author | TP | TC | CPP | h-Index | Self-Citations |
|---|---|---|---|---|---|---|
| 1 | Akkaya, K. | 11 | 74 | 6.73 | 4 | 6 |
| 2 | Erdin, E. | 9 | 34 | 3.78 | 4 | 6 |
| 3 | Yang, D. | 8 | 34 | 4.25 | 3 | 6 |
| 4 | Javaid, N. | 7 | 48 | 6.86 | 4 | 15 |
| 5 | Kumar, N. | 7 | 125 | 17.86 | 5 | 49 |
| 6 | Salah, K. | 7 | 553 | 79.00 | 6 | 45 |
| 7 | Zhang, C. | 7 | 13 | 1.86 | 2 | 1 |
| 8 | Cebe, M. | 6 | 57 | 9.5 | 4 | 4 |
| 9 | Kanhere, S.S. | 6 | 82 | 13.67 | 3 | 13 |
| 10 | Liyanage, M. | 6 | 127 | 21.17 | 5 | 14 |

Note(s): This table was created with a dataset from Scopus

The authors with the most impactful publications, on average, as measured by CPP, are Salah, K., Liyanage, M., and Kumar, N. However, out of Kumar, N.'s 125 total citations, 39.2% were self-citations. The most productive author in this group is Salah, K., who has an h-Index of 6. Notably, Salah, K. also has a CPP of 79.00. Akkaya, K. has authored the most publications on blockchain-based payments. Most of his contributions to the research space lie in improving the transaction success rate in cryptocurrency payments and preserving privacy in IoT Micro-payments (Kurt et al. 2022; Mercan et al. 2019).

Contrary to Table 4, which shows citations based on total publications, Table 5 shows the top 10 individual studies with the most citations. The author and publishing year are listed to operationalize each document. The year column lists the number of citations received during that period. In the % column, the percent change between the preceding and the following year is identified between each year. The total number of citations received for these 10 papers is 1,782, with an average of 178.2 citations per academic document.

**Table 5.** Top 10 most cited blockchain-based payments publications

| No. | Publications/Year | 2018 | % | 2019 | % | 2020 | % | 2021 | % | 2022 | % | Total |
|---|---|---|---|---|---|---|---|---|---|---|---|---|
| 1 | Salah et al. (2019) | 0 | 0% | 48 | 14% | 90 | 27% | 112 | 33% | 85 | 25% | 335 |
| 2 | Kamble et al. (2020) | 0 | 0% | 0 | 0% | 40 | 15% | 105 | 40% | 120 | 45% | 265 |
| 3 | De' et al. (2020) | 0 | 0% | 0 | 0% | 15 | 6% | 89 | 36% | 140 | 57% | 244 |
| 4 | Li et al. (2019) | 0 | 0% | 6 | 4% | 31 | 19% | 70 | 42% | 60 | 36% | 167 |
| 5 | Thakor (2020) | 0 | 0% | 0 | 0% | 20 | 13% | 53 | 35% | 79 | 52% | 152 |
| 6 | Wang and Wang (2019) | 0 | 0% | 4 | 3% | 43 | 30% | 59 | 41% | 39 | 27% | 145 |
| 7 | Alladi et al. (2019) | 0 | 0% | 0 | 0% | 29 | 21% | 58 | 42% | 52 | 37% | 139 |
| 8 | Milian et al. (2019) | 0 | 0% | 4 | 3% | 27 | 22% | 44 | 36% | 46 | 38% | 121 |
| 9 | Alladi et al. (2019) | 0 | 0% | 0 | 0% | 28 | 26% | 51 | 47% | 30 | 28% | 109 |
| 10 | Lao et al. (2020) | 0 | 0% | 0 | 0% | 20 | 19% | 53 | 50% | 32 | 30% | 105 |

Note(s): This table was created with a dataset from Scopus

*4.6 Documents by Type*

Figure 4 shows the percent distribution for the different ways authors present their research (about blockchain-based payments).

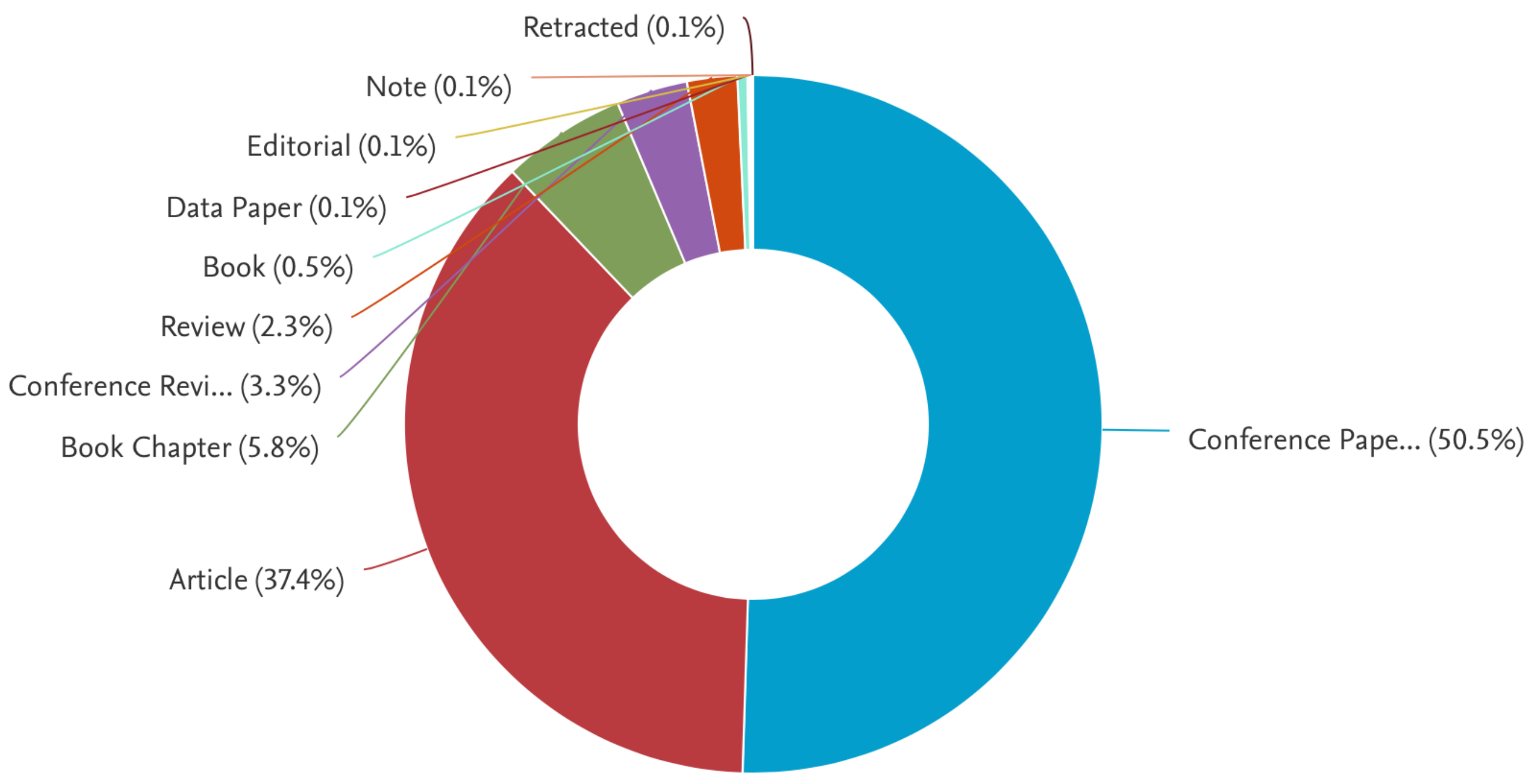

**Figure 4.** These are the percentages for the different document types of educational documents (in the final stage)

concerning blockchain-based payments between 2019 and 2022. The data and pie-chart were retrieved through the Scopus database.

The quantified data is as follows, 762 conference papers, 566 articles, 87 book chapters, 50 conference reviews, 35 reviews, 7 books, 1 data paper, 1 editorial, 1 note, and 1 retracted piece.

*4.7 Distribution of Disciplines*

Figure 5 shows the segmentation of subject areas for literature in the blockchain-based payment space between 2018 and 2022.

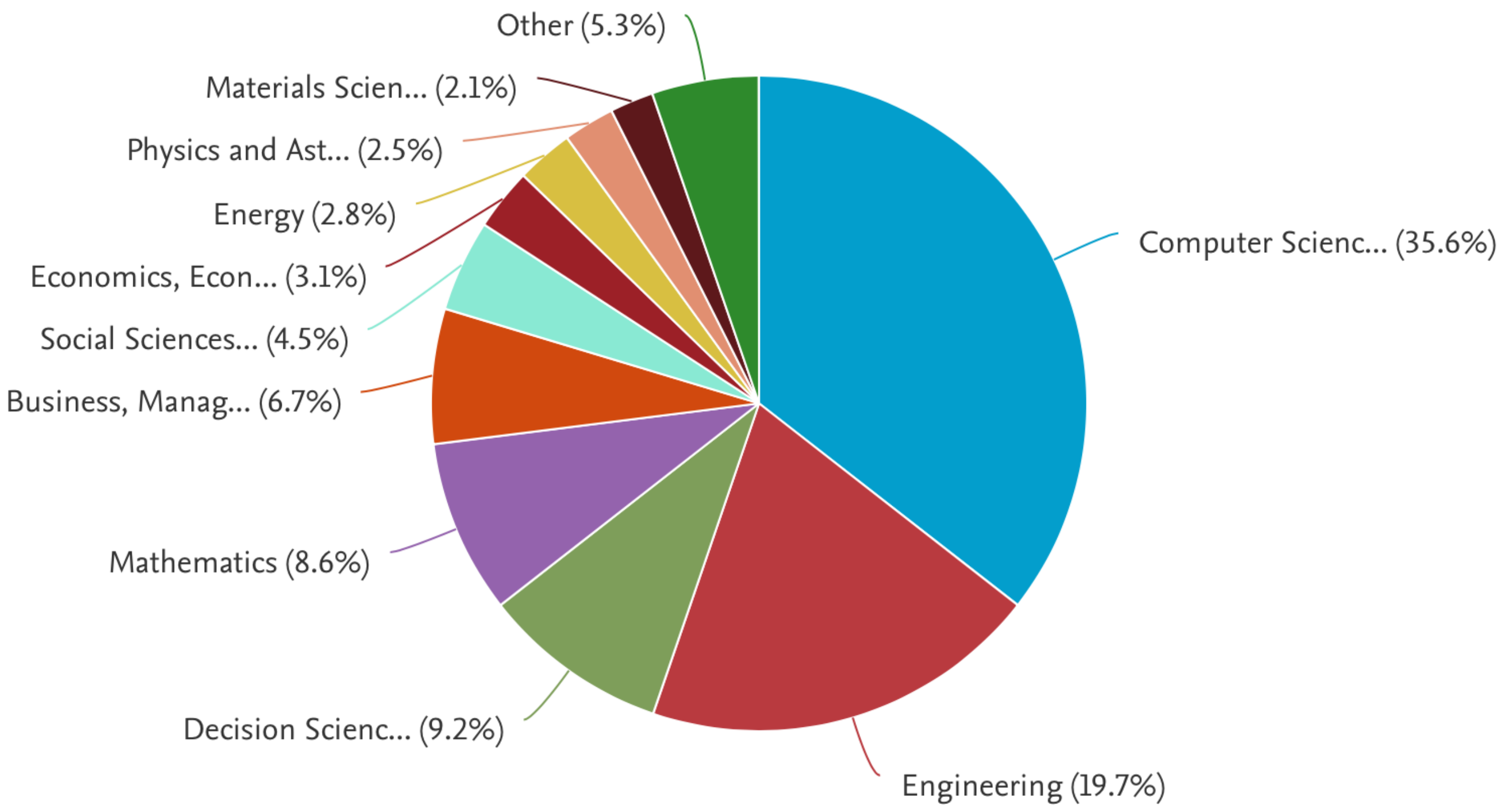

**Figure 5.** These are the percentages for the different subject areas of educational documents (in the final stage) concerning blockchain-based payments between 2019 and 2022. The data and pie-chart were retrieved through the Scopus database.

The quantified data can be identified by multiplying the percentages by 1,511, which was the sample size. There are 11 subject areas in the "Other" section: environmental science, medicine, chemical engineering, chemistry, Earth and planetary sciences; multidisciplinary, biochemistry, genetics and molecular biology; arts and humanities; agricultural and biological sciences; and psychology. These disciplines were listed in descending order of the number of documents published.

*4.8 Keyword Co-occurrence Analysis*

To conduct a keyword co-occurrence analysis, VOSviewer extracts keywords from abstracts and titles and runs a clustering algorithm to create accurate networks (van Eck and Waltman, 2011, as cited in Moosavi et al. 2021). The process began by exporting the bibliographic data from Scopus to a CSV file in VOSviewer ("Scopus database & VOSViewer: Extracting data from Scopus for Bibliometric & scientometric analyses." 2020). VOSviewer offers 3 units of analyses under Co-occurrence which are "all keywords," "author keywords," and "index keywords." This study selected "all keywords" to synthesize all co-occurrence data. Furthermore, a full counting method gave each co-occurrence link the same weight. The keyword must have been mentioned at least 5 times to maintain relevance. Of the original 7,572 keywords, only 540 met this threshold. This resulted in 9 unique clusters, with 17,242 links and a total link strength of 35,632. Figure 6 presents a graphical model of the aforementioned. Table 6 illustrates the number of times the 14 most common keywords were mentioned. The keywords are listed in descending order of link strength.

**Table 6.** The most repeated keywords and their number of occurrences

| Keywords | No. of occurrences |
|---|---|
| Blockchain | 1204 |
| Block-chain | 304 |
| Smart Contract | 232 |
| Bitcoin | 228 |
| Cryptocurrency | 182 |
| Internet of Things | 133 |
| Cryptography | 125 |
| Network Security | 104 |
| Ethereum | 127 |
| Electronic Money | 115 |
| Peer to Peer Networks | 90 |
| Smart Contracts | 108 |
| Payment Systems | 89 |
| Digital Storage | 77 |

Note(s): This table was created with datasets from Scopus and VOSviewer

The most repeated keywords in these papers are blockchain, smart contracts, and Bitcoin. There is a low correlation between blockchain and payment systems as “payment systems” are only mentioned 89 times while “blockchain” is mentioned 1204 times.

Figure 6 shows the network of keywords. The size of each node is relative to the number of times it was mentioned in publications. The lines connecting the nodes represent the relationships between these keywords. The 9 different colors constitute 9 different clusters.

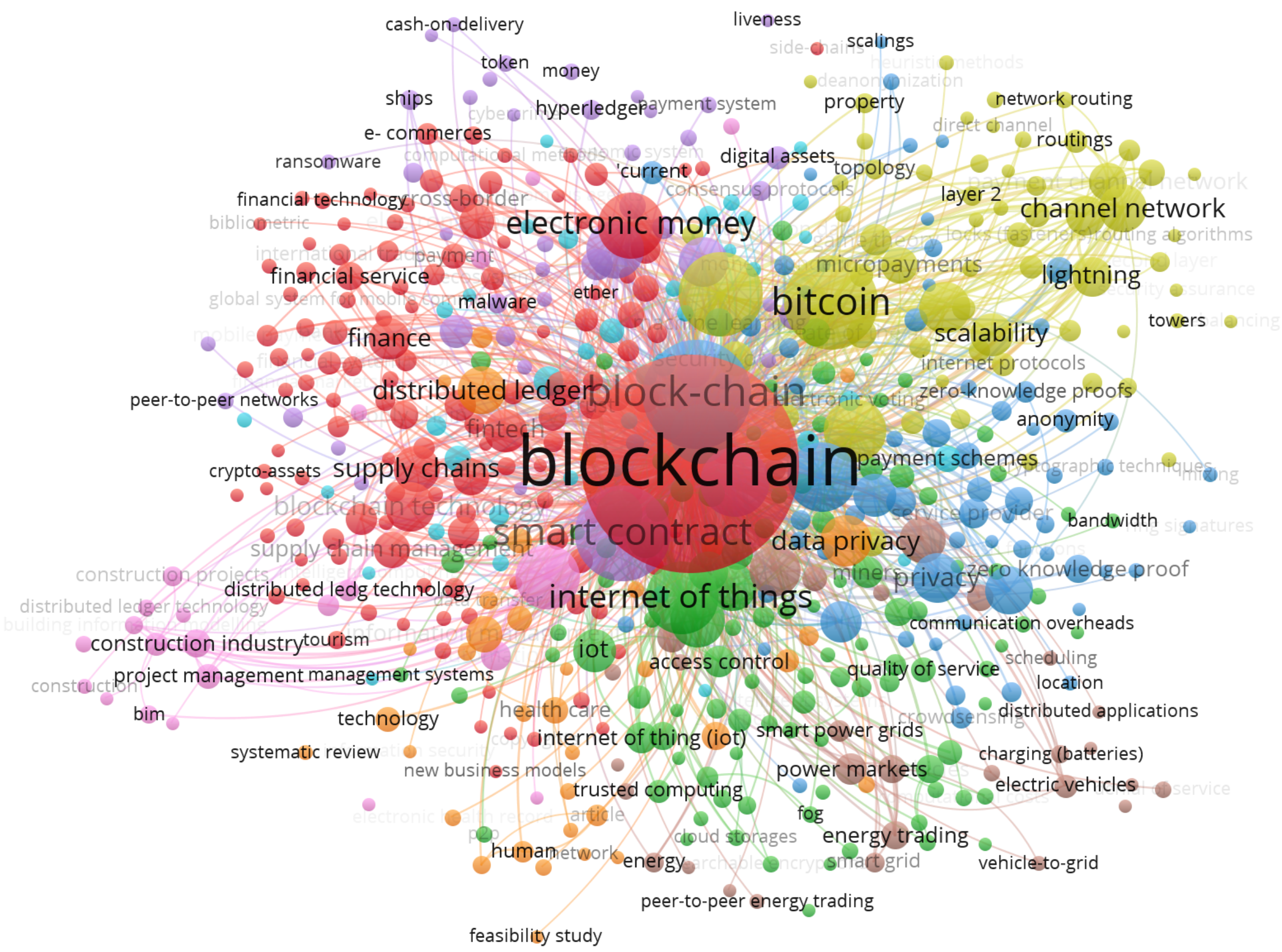

**Figure 6.** Keywords and their connection network map. The dataset was gathered through Scopus, and the network model was created through VOSviewer.

### *4.9 Co-authorship Networks*

Co-authorship analysis has many implications for understanding research collaboration and identifying its patterns. Collaboration is imperative to discovering knowledge as it allows for the efficient fulfillment of tasks (Sonnenwald 2008, as cited in Zicker 2016). Furthermore, when researchers mutually share goals, innovation and productivity increase exponentially. The combination of different skills and knowledge allows researchers to broaden the scope of their projects, increasing their impact (Deb 2001, as cited in Zicker 2016).

There are many applications for co-authorship analysis. By and large, it helps to create bridges between the research community. Identifying the most active contributors in a discipline makes associating groups best suited to lead a project easier.

This paper's co-authorship analysis will assess inter-organizational networks, international collaboration, and regional contributions to knowledge generation.

In co-authorship networks, nodes represent authors, organizations, or countries (depending on the context), which are connected when they share the authorship of the paper. This allows for graphical visualization of research collaboration (Deb 2001, as cited in Zicker 2016). VOSviewer provides three types of visualization which are "network visualization," "overly visualization," and "density visualization." The graphs in this study all use network visualization.

Using the CSV file downloaded from Scopus, VOSviwer was granted access to this study's data for this study.

*4.91 Co-authorship Network of Authors*

VOSviewer offers 3 units of analyses under co-authorship, which are “authors,” “organizations,” and “countries.” Section 4.91 selected “authors.” Furthermore, a full counting method was chosen, giving each co-authorship link the same weight. To ensure that this study only mapped the most prominent authors in the field, they must have at least 10 published documents and 50 citations. Of the original 3,692 authors, only 17 met this threshold. However, some of the 17 authors were not connected. Since section 4.9 attempts to visualize relationships, all items that were not connected were omitted to leave 16 authors. This resulted in 4 unique clusters, with 29 links and a total link strength of 37. Figure 7 presents a graphical model of the aforementioned. Table 7 illustrates the top research groups of authors contributing to blockchain in payment systems.

**Table 7.** The top research groups of authors contributing to blockchain-based payments literature

| Cluster 1 | | |
|---|---|---|
| Author | No. of docs | No. of links |
| Chen, Y. | 10 | 4 |
| Li, J. | 13 | 2 |
| Wang, Y. | 11 | 4 |
| Zhang, J. | 12 | 3 |
| Zhang X. | 14 | 3 |

| Cluster 2 | | |
|---|---|---|
| Author | No. of docs | No. of links |
| Chen, J. | 12 | 6 |
| Liu, Z. | 11 | 3 |
| Wang, H. | 12 | 1 |
| Zhang Y. | 20 | 6 |

| Cluster 3 | | |
|---|---|---|
| Author | No. of docs | No. of links |
| Li, Y. | 17 | 2 |
| Wang, Q. | 10 | 4 |
| Yang, Z. | 10 | 5 |

| Cluster 4 | | |
|---|---|---|
| Author | No. of docs | No. of links |
| Liu, J. | 10 | 2 |
| Sun, Y. | 11 | 4 |
| Wang, S. | 11 | 6 |

Note(s): This table was created with datasets from Scopus and VOSviewer

The group with the most publications is Cluster 1 as they have authored 60 documents together. On the other hand, Cluster 2 has the most connections at 16 links.

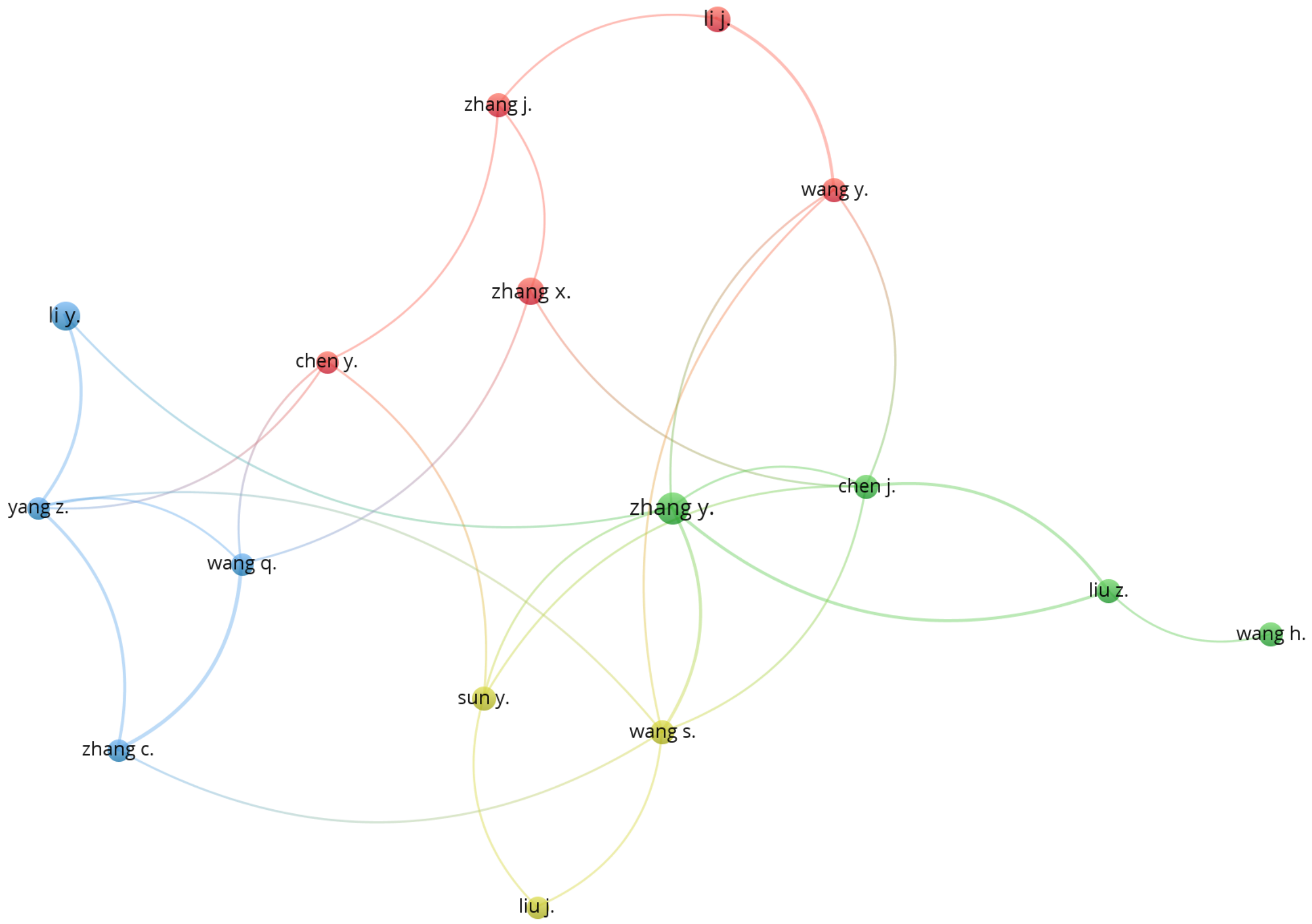

**Figure 7.** A network model representing co-authorship between researchers on educational documents concerning blockchain-based payments between 2019 and 2022. This data was retrieved through the Scopus database, and the graph was created through VOSviewer.

*4.92 Co-authorship Network of Organizations*

VOSviewer offers 3 units of analyses under co-authorship, which are "authors," "organizations," and "countries." Section 4.92 selected "organizations." Furthermore, a full counting method was chosen, giving each co-authorship link the same weight.  To ensure that this study only mapped the most prominent organizations in the field, they must have at least 1 published document and 10 citations. Of the original 2,880 organizations, 669 met this threshold. However, some of the 669 organizations were not connected. Since section 4.9 attempts to visualize relationships, all items that were not connected were omitted to leave 15 organizations. This resulted in 3 unique clusters, with 47 links and a total link strength of 48. Figure 8 presents a graphical model of the aforementioned. Table 8 illustrates the top research groups of organizations contributing to blockchain in payment systems.

**Table 8.** The top research groups of organizations contributing to blockchain-based payments literature

| Cluster 1 | | |
|---|---|---|
| Organization | No. of docs | No. of links |
| Department of Information Systems, Cybersecurity and the Department of Electrical and Computer Engineering, the University of Texas at San Antonio | 1 | 6 |
| Fujian Provincial Key Laboratory of Network Security and Cryptology, Center for Applied Mathematics of Fujian Province, College of Mathematics and Informatics, Fujian Normal Uni. | 1 | 6 |
| Shandong Provincial Key Laboratory of Computer Networks, Qilu University of Technology | 1 | 6 |
| Shanghai Key Laboratory of Privacy-Preserving Computation, Matrixelements Technologies | 1 | 6 |
| State Key Laboratory of Cryptology | 2 | 12 |
| State Key Laboratory of Information Security, Institute of Information Engineering, Chinese Academy of Sciences | 4 | 6 |

| Cluster 2 | | |
|---|---|---|
| Organization | No. of docs | No. of links |
| Center of Excellence in Information Assurance (Coeia), King Saud University | 1 | 6 |
| College of Mathematics and Informatics, Fujian Normal University | 1 | 6 |
| Cyberspace Security Research Center, Peng Cheng Laboratory | 2 | 6 |
| Department of Information Systems and CyberSecurity, the University of Texas at San Antonio | 1 | 6 |
| Key Laboratory of Aerospace Information Security and Trusted Computing, Ministry of Education, School of Cyber Science and Engineering, Wuhan University | 1 | 6 |

| Cluster 3 | | |
|---|---|---|
| Organization | No. of docs | No. of links |
| Computer Science Department, City University of Hong Kong | 1 | 3 |
| School of Computer Science, Wuhan University | 1 | 3 |
| School of Cyber Science and Engineering, Wuhan University | 4 | 14 |
| State Key Laboratory of Mathematical Engineering and | 1 | 3 |
| Advanced Computing | | |

Note(s): This table was created with datasets from Scopus and VOSviewer

The group with the most publications is Cluster 1 as they have authored 10 documents together. On the other hand, Cluster 2 has the most connections at 42 links.

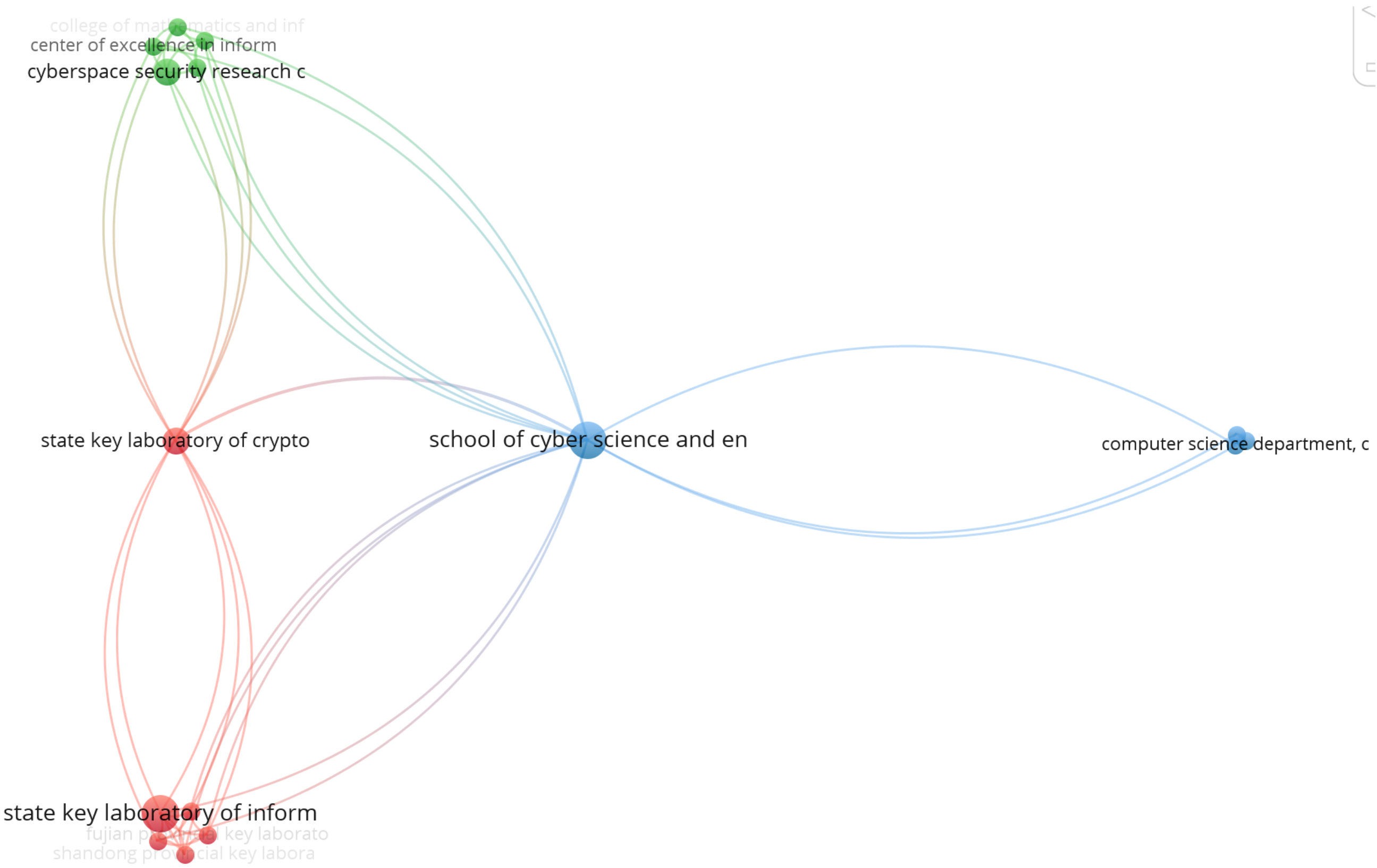

**Figure 8.** A network model representing co-authorship between organizations on educational documents concerning blockchain-based payments between 2019 and 2022. This data was retrieved through the Scopus database, and the graph was created through VOSviewer.

*4.93 Co-authorship Network of Countries*

VOSviewer offers 3 units of analyses under co-authorship, which are “authors,” “organizations,” and “countries.” Section 4.93 selected “countries.” Furthermore, a full counting method was chosen, giving each co-authorship link the same weight.  To ensure that this study was only mapping the most prominent countries in the field, they must have at least published 5 documents together and have 100 citations. Of the original 132 countries, 34 met this threshold. However, some of the 34 countries were not connected. This resulted in 4 unique clusters, with 218 links and a total link strength of 549. Figure 9 presents a graphical model of the aforementioned. Table 9 illustrates the top research groups of countries contributing to blockchain in payment systems.

**Table 9.** The top research groups of countries contributing to blockchain-based payments literature

| Cluster 1 | | | Cluster 2 | | |
|---|---|---|---|---|---|
| Country | No. of docs | No. of links | Country | No. of docs | No. of links |
| Brazil | 16 | 9 | Austria | 27 | 13 |
| Canada | 59 | 18 | Germany | 57 | 13 |
| France | 35 | 17 | Israel | 15 | 5 |
| India | 204 | 22 | Italy | 47 | 14 |
| Pakistan | 24 | 14 | Netherlands | 20 | 13 |
| Portugal | 12 | 11 | Poland | 10 | 5 |
| Qatar | 17 | 10 | Russian Federation | 45 | 10 |
| Saudi Arabia | 36 | 13 | Singapore | 27 | 14 |
| South Korea | 44 | 12 | Spain | 32 | 16 |
| Taiwan | 26 | 12 | Switzerland | 39 | 15 |
| United States | 208 | 28 | Việt Nam | 12 | 8 |

| Cluster 3 | | | Cluster 4 | | |
|---|---|---|---|---|---|
| Country | No. of docs | No. of links | Country | No. of docs | No. of links |
| Australia | 77 | 24 | China | 347 | 25 |
| Belgium | 7 | 6 | Hong Kong | 40 | 8 |
| Finland | 13 | 8 | Japan | 28 | 10 |
| Greece | 18 | 6 | Jordan | 9 | 7 |
| Ireland | 12 | 10 | Norway | 8 | 4 |
| United Kingdom | 84 | 27 | United Arab Emirates | 35 | 9 |

Note(s): This table was created with datasets from Scopus and VOSviewer.

The group with the most publications is Cluster 1, as they have authored 681 documents together. On the other hand, Cluster 2 has the most connections at 166 links. A notable characteristic of this table is that these clusters consider geographical proximity. For example, out of Cluster 3's 6 countries, 5 are from Europe.

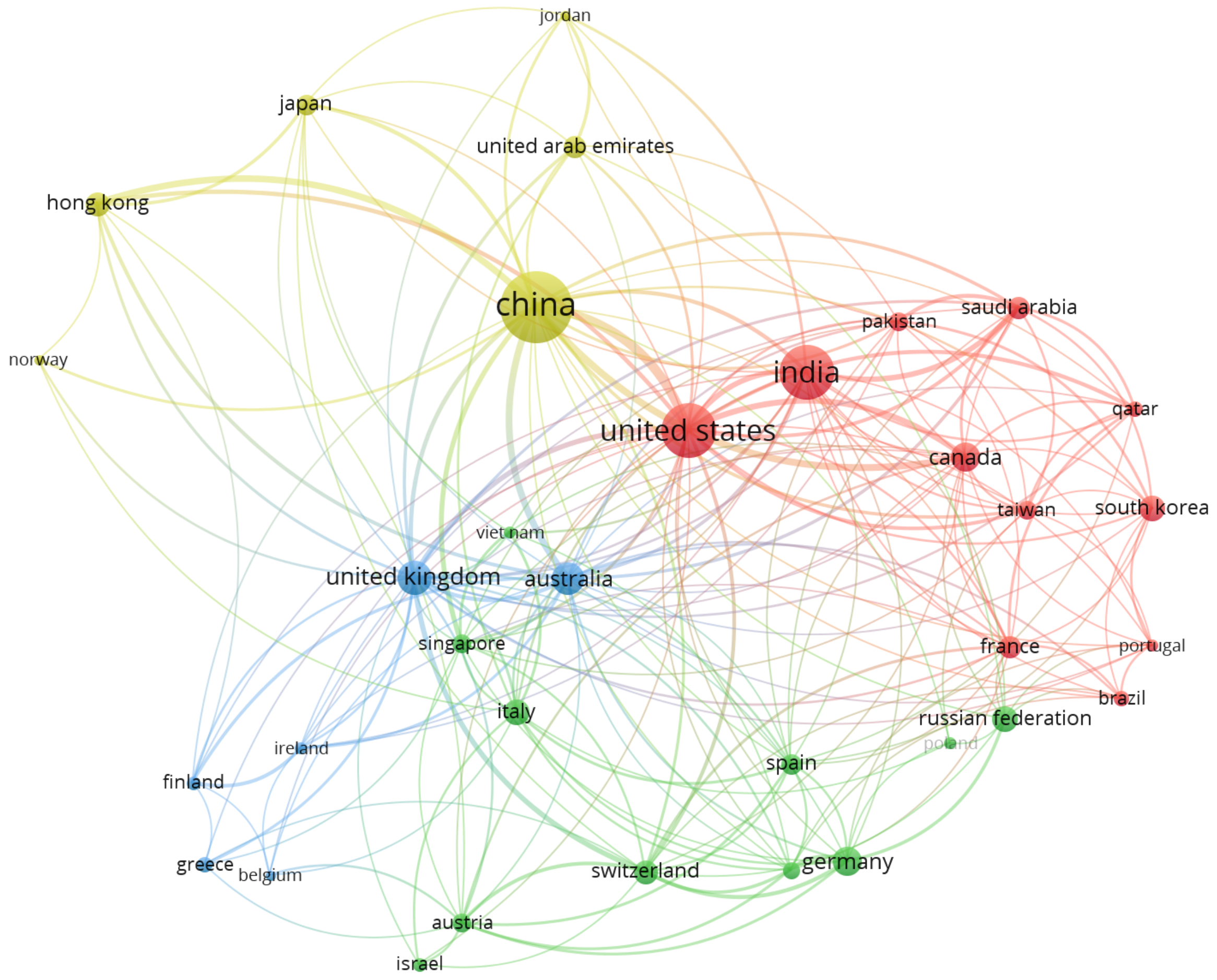

**Figure 9.** A network model representing co-authorship between countries on educational documents concerning blockchain-based payments between 2019 and 2022. This data was retrieved through the Scopus database, and the graph was created through VOSviewer.

## 5. Results and Discussion

RQ1. How has literature on blockchain-based payment systems developed between 2019 and 2022?

This question will be answered by subquestions RQ1.1, RQ1.2, RQ1.3, and RQ1.4.

RQ1.1. What are the most influential studies, affiliations, and authors?

Section 4.5, labeled author influence, identifies the top 10 contributing authors to blockchain-based literature. This study operationalizes the variable "influence" by CPP (to measure impact) and h-Index (to measure productivity). With this in mind, the top three authors are Salah, K., Liyanage, M., and Kumar, N. Salah K. has an h-index of 6 and over 553 TC making his CPP (79.00) the highest in this paper. Salah, K. is a published researcher for the Khalifa Institute for Science and Technology. He has over 220 publications and is a leader in blockchain, IoT, cybersecurity, and cloud computing. Salah's current research focuses on using blockchain to alleviate pressure on the supply chain and blockchain's applications in healthcare (Salah). However, his past work also indicates an interest in blockchain-based payment systems (Hasan and Salah 2018).

Table 5 looks at the most cited publications for blockchain-based payments. After adding up the TC received per year, Salah et al. (2019) were the highest at 335. This paper was titled “Blockchain for AI: Review and Open Research Challenges.” It presents a detailed description of the many applications of blockchain in AI. However, it also addresses remittances and financial payments. Concerning payment, Salah also elaborates on Ethereum and its ability to automate transactions (Salah et al. 2019). To analyze document type, Figure 4 shows that most of these documents are conference papers. The second most prevalent format is articles. Together, they make up 87.9% of all publications.

Table 5 also shows the distribution of TC over the years 2018-2022. The data from table 5 allows for the observation of general trends. In 2018, researchers did not author academic papers on the topic, and in 2019 growth was slow. In 2020 researcher interest began to increase; however, the true exponential growth occurred in 2021. Every publication received at least 30% of its citations during this time. In 2022, a slight plateau was observed, and despite some papers like (Kamble et al. 2020) receiving 57% of their citations, papers like (Saleh et al. 2019) were only at a 25% increase. This more expansive range in recent years could represent that the research space is more diverse, with many specializations.

Lastly, Table 1 analyzes the top 10 affiliations contributing to blockchain-based payments literature. The most influential affiliation is the Khalifa University of Science and Technology. This shows a correlation between author and affiliation influence, as Salah, K, is a professor at this university (Salah). With this establishment based in the United Arab Emirates, it is 1 of 6 affiliations in Asia. The unique geographical proportions express both Asian interest and influence in the field.

RQ1.2. What are the geographical distributions and effects of publications over time?

Despite Asian countries having the most affiliations in the top 10, the United States has the highest CPP and h-Index. They have the highest h-Index observed in this study at 25, representing a high productivity rate. China has 139 more publications than the US, but since their citations did not rise proportionately (and are below the US's by 415), their CPP is half that of the US. Notably, the United Kingdom has the second-highest CPP, at 7.76. Their high CPPs could explain the US-UK collaboration seen in Figure 9. Identifying geographical distributions has many implications for the future of blockchain-based payment systems. The global community maximizes efficiency by pinpointing the regions most suited to tackle a research issue. Figure 2 graphically illustrates the publication trend by year. Despite a gradual incline from 2019-2021, there was a sharp decline of 43.86% between 2021 and 2022. This decreasing interest in 2022 can be attributed to the low rates of blockchain adoption (De Meijer 2020). The general public has associated the Bitcoin crash with blockchain, and thus, interest in it has decreased. If blockchain is to optimize day-to-day operations forever, then the quantity of publications needs to bounce back, continuing the trend from 2018-2021.

RQ1.3. Which funding sponsors have the most extensive contributions?

Table 3 shows the top 10 funding sponsors that contributed to blockchain-based payments. The National Science Foundation of China funds significantly more papers than other organizations. Their TP is 3x higher than the next best sponsor, the National Science Foundation. The NSFC also has an h-Index double that of the next best alternative. This demonstrates that their papers are productive in their research methods. Despite this, they do not have the highest CPP, as the Fundamental Research Funds for the Central Universities overtook them by 2.47. The NSFC is committed to funding blockchain solutions, which helped blockchain research develop during 2019-2022. Their perseverance is seen through their attempts to use blockchain for health records during the COVID-19 pandemic (Odoom et al., 2022). Moving forward, the NSFC is a great option for funding blockchain research. This is corroborated by their budget for research and related activities which makes up 81% of the funding ($73.4 billion) they provide ("Funding and Support Descriptions | NSF").

RQ2. What are the most important topics in blockchain-based payments literature?

Two parts of the study must be referenced to answer this question.

First is section 4.7, or the distribution of disciplines. This pie chart shows the segmentation of subject areas for literature in the space. Just by observing this graph, it is clear that blockchain has countless applications. Out of the 22 fields of study, the top 5 are computer science, engineering, decision science, mathematics, and business. Together they make up 79.8% of all publications. Despite not being categorized, blockchain payments can fall into the labels of computer science (smart contracts) and business (remittances and trade finance). When looking at blockchain's use cases in each of these fields, its primary purpose becomes more clear (Kumar 2018; Viswanathan 2022; Gibbons 2019). These 5 sectors use blockchain to record and store data, preferring it due to its transparency and immutability.

Second is section 4.8, where a keyword co-occurrence analysis was carried out. After VOSviewer scrutinized 1511 abstracts and titles, a list of keywords was created. This list included the number of occurrences that each keyword was mentioned. The most repeated are "blockchain," "smart contract," "Bitcoin," and "cryptocurrency." These keywords strongly support the future of blockchain-based payments, even though "payment systems" are not directly mentioned. This is because smart contracts are primarily used to automate monetary affairs ("Using Smart Contracts for Automated Financial Features, Like Debt Repayment • Sila"). In contrast, Bitcoin and cryptocurrency use blockchain as a digitalized ledger to store transaction data (Little 2022). In summary, as seen through the keywords, researchers rightly emphasize the importance of blockchain-based payments.

RQ3. What are the relationships among authors, countries, and organizations in blockchain-based payments literature?

Section 4.91 identified 4 clusters (research groups) of authors. These groups were created based on TP, TC, and whether they have co-authored a publication. The authors with the most links were Chen, J., Zhang, Y., and Wang, S. The many links emanating off their nodes hold the network together. Intriguingly, the authors in each cluster share backgrounds. For example, in Cluster 4, all authors had cryptography and computer science experience. However, by combining different expertise, researchers would be able to expand the breadth of their research. When this fundamental change is made, research groups will become more impactful and productive ("5 ways that collaboration can further your research and your career | For Researchers" 2019).

On the other hand, for organizations, 3 unique clusters were created in Table 8. Notably, the School for Cyber Science and Engineering and the State Key Laboratory of Cryptography were bridges between the organizations graphed in Figure 8. This can be seen visually and through their large number of links–-14 and 12, respectively, compared to the mean of 6.33 links. Interestingly, intra-organizational cooperation is not prevalent in blockchain-based payment research. The maximum number of documents co-authored is only 4. Instead of institutions competing, they should cooperate for mutually shared goals. This would increase efficiency and capitalize on knowledge specializations (Bansal et al. 2019).

Lastly, for countries, 4 unique clusters were created in Table 9. The countries with the most links were the United States, the United Kingdom, and China. By analyzing the research groups formed in Table 9, it is clear that geographical location impacts research collaboration. For example, in Cluster 3, 5 of 6 countries were from Europe. In Cluster 4, 4 of 6 countries were from Asia. Thus, it is important to encourage national research collaboration, as it increases efficiency and bridges disciplines ("Research collaborations bring big rewards: the world needs more" 2021).

## 5. Conclusion and Future Research Direction

Although some articles have been published regarding blockchain's implications in payment, bibliometric or network analyses are scarce. This study contributes to the current understanding of blockchain-based payments in three different ways in the scope of 1,511 publications obtained from Scopus between 2019 and 2022:

- This study analyzes the development of literature over 4 years of research in 6 key ways: summarizing the publication trend, identifying university affiliations, studying geographical distributions, recognizing extensive funding sponsors, distinguishing influential authors, and determining the distributions between types of documents.
- This paper establishes the most important sub-fields in blockchain-based payments literature through an author and index keyword co-occurrence analysis and by determining the distributions of disciplines.
- This study creates global, organizational, and authorial networks to chart collaborations between individual nodes.

One limitation of this study is that it searched academic papers solely from Scopus instead of diversifying through the Web of Science and ScienceDirect databases. It also used a two-level keyword search when selecting documents, omitting some relevant publications.

In conclusion, the main future research directions can be pointed out as follows:

- Researching the correlation between blockchain-based payments and the subfields of energy, decision sciences, the economy, and mathematics.
- Conducting bibliometric analyses per organization/country to identify research strengths and inspire efficiency through networking strategic research partners.
- Identifying methods to scale blockchain-based payments while preserving the inherent security of each transaction.

## 6. References

Alladi, T., et al. "Blockchain Applications for Industry 4.0 and Industrial IoT: A Review." *IEEE Access*, vol. 7, 2019, pp. 176935-176951. *SCOPUS*, www.scopus.com, doi:10.1109/ACCESS.2019.2956748.

Alladi, T., et al. "Blockchain in Smart Grids: A Review on Different use Cases." *Sensors (Switzerland)*, vol. 19, no. 22, 2019. *SCOPUS*, www.scopus.com, doi:10.3390/s19224862.

Alvesson, Mats, and Jorgen Sandberg. *Constructing Research Questions: Doing Interesting Research*. SAGE Publications, 2013.

Bansal, Seema, et al. "Collaborative research in modern era: Need and challenges." *NCBI*, 27 June 2019, https://www.ncbi.nlm.nih.gov/pmc/articles/PMC6644188/. Accessed 7 October 2022.

"BeckerGuides: Tools for Authors: What is the h index?" *BeckerGuides at Becker Medical Library*, 2 September 2022, https://beckerguides.wustl.edu/authors/hindex. Accessed 3 October 2022.

"Blockchain in Payments | Blockchain Use Cases." *LeewayHertz*, https://www.leewayhertz.com/blockchain-in-payments/. Accessed 6 October 2022.

Cryptovision. "Avalanche effect." *cv cryptovision GmbH*, 15 April 2021, https://www.cryptovision.com/en/glossary/avalanche-effect/. Accessed 5 October 2022.

Darabseh, M., and JP Martins. "Risks and Opportunities for Reforming Construction with Blockchain: Bibliometric Study." *Repositório Aberto da Universidade do Porto*, 2020, https://repositorio-aberto.up.pt/handle/10216/127468. Accessed 6 October 2022.

Deb, Donald. "Reflections on Scientific Collaboration (and its study): Past, Present, and Future - Scientometrics." *Springer*, 2001, https://link.springer.com/article/10.1023/A:1014254214337. Accessed 4 October 2022.

De Meijer, Carlo RW. "Remaining challenges of blockchain adoption and possible solutions." *Finextra*, 29 February 2020, https://www.finextra.com/blogposting/18496/remaining-challenges-of-blockchain-adoption-and-possible-solutions. Accessed 7 October 2022.

De', R., N. Pandey, and A. Pal. "Impact of Digital Surge during Covid-19 Pandemic: A Viewpoint on Research and Practice." *International Journal of Information Management*, vol. 55, 2020. *SCOPUS*, www.scopus.com, doi:10.1016/j.ijinfomgt.2020.102171.

Donthu, Naveen, et al. "How to conduct a bibliometric analysis: An overview and guidelines." *ScienceDirect*, 14 May 2021, https://www.sciencedirect.com/science/article/pii/S0148296321003155. Accessed 2 September 2022.

Duan, Ruijun, and Li Guo. "Application of Blockchain for Internet of Things: A Bibliometric Analysis." *Hindawi*, 29 April 2021, https://www.hindawi.com/journals/mpe/2021/5547530/. Accessed 6 October 2022.

Elsaid, Haitham. "The application of blockchain in trade finance: opportunities and challenges." *Trade Finance Global*, 17 November 2020, https://www.tradefinanceglobal.com/posts/the-application-of-blockchain-in-trade-finance-opportunities-and-challenges/. Accessed 6 October 2022.

"5 ways that collaboration can further your research and your career | For Researchers." *Springer Nature*, 11 November 2019, https://www.springernature.com/gp/researchers/the-source/blog/blogposts-life-in-research/benefits-of-research-collaboration/17360752. Accessed 7 October 2022.

"Funding and Support Descriptions | NSF." *National Science Foundation*, https://www.nsf.gov/homepagefundingandsupport.jsp#where-it-comes-from. Accessed 7 October 2022.

Gao, Yi-Ming, et al. "A bibliometric analysis and visualization of blockchain." *Future Generation Computer Systems*, vol. 116, 2021, pp. 316-332. *ScienceDirect*, https://www.sciencedirect.com/science/article/pii/S0167739X20330004?casa_token=sUwC_aY-h7kAAAAA:HTVZgIBih4YyyYX7R5r9MAl40neFLOrNlPiKmKzX5100Yme0DZo-SmuG9wYbjjgGzkI7vPTcCw. Accessed 6 October 2022.

Gibbons, Serenity. "3 Practical Ways To Use Blockchain In Your Business In 2020." *Forbes*, 19 November 2019, https://www.forbes.com/sites/serenitygibbons/2019/11/19/3-practical-ways-to-use-blockchain-in-your-

business-in-2020/?sh=6fb7e2372b7b. Accessed 7 October 2022.

Hasan, Haya R., and Khaled Salah. “Proof of Delivery of Digital Assets Using Blockchain and Smart Contracts.” *IEEE Access2*, 22 October 2018, https://ieeexplore.ieee.org/stamp/stamp.jsp?arnumber=8501910. Accessed 7 October 2022.

Hayes, Adam. “Blockchain Facts: What Is It, How It Works, and How It Can Be Used.” *Investopedia*, 27 September 2022, https://www.investopedia.com/terms/b/blockchain.asp. Accessed 5 October 2022.

Hayes, Adam. “Who Is Satoshi Nakamoto?” *Investopedia*, 25 September 2022, https://www.investopedia.com/terms/s/satoshi-nakamoto.asp. Accessed 5 October 2022.

Hirsch, J. E. “An index to quantify an individual's scientific research output.” *PNAS*, 7 November 2005, https://www.pnas.org/doi/10.1073/pnas.0507655102. Accessed 3 October 2022.

Iredale, Gwyneth. “Blockchain Technology History: Ultimate Guide.” *101 Blockchains*, 3 November 2020, https://101blockchains.com/history-of-blockchain-timeline/. Accessed 5 October 2022.

Kamran, Muhammad, et al. “Blockchain and Internet of Things: A bibliometric study.” *Computers & Electrical Engineering*, vol. 81, 2020. *ScienceDirect*, https://www.sciencedirect.com/science/article/pii/S0045790618333913?casa_token=WZwmrlOSmaUAAAAA:M1igyEqMzfCSVo8k5txaNEFnX1pR2WE90hcbBYXKRLodCo0m5_tkWsJIC27XtVQx9HINms2l4Q. Accessed 6 October 2022.

Kamble, S. S., A. Gunasekaran, and R. Sharma. "Modeling the Blockchain Enabled Traceability in Agriculture Supply Chain." *International Journal of Information Management*, vol. 52, 2020. *SCOPUS*, www.scopus.com, doi:10.1016/j.ijinfomgt.2019.05.023.

Klebanov, Sam. “A Look at Blockchain in Cross-Border Payments.” *PaymentsJournal*, 5 February 2021, https://www.paymentsjournal.com/a-look-at-blockchain-in-cross-border-payments/. Accessed 6 October 2022.

Kumar, Sree. “Blockchain technology and Digital engineering.” *ResearchGate*, June 2018, https://www.researchgate.net/publication/325677392_Blockchain_technology_and_Digital_engineering. Accessed 7 October 2022.

Kurt, Ahmet, et al. “LNGate2: Secure Bidirectional IoT Micro-payments using Bitcoin’s Lightning Network and Threshold Cryptography.” *arXiv*, 17 January 2022, https://arxiv.org/pdf/2206.02248.pdf. Accessed 10 November 2022.

Lao, L., et al. "A Survey of IoT Applications in Blockchain Systems: Architecture, Consensus, and Traffic Modeling." *ACM Computing Surveys*, vol. 53, no. 1, 2020. *SCOPUS*, www.scopus.com, doi:10.1145/3372136.

Li, J., D. Greenwood, and M. Kassem. "Blockchain in the Built Environment and Construction Industry: A Systematic Review, Conceptual Models and Practical use Cases." *Automation in Construction*, vol. 102, 2019, pp. 288-307. *SCOPUS*, www.scopus.com, doi:10.1016/j.autcon.2019.02.005.

Li, Zhi, and Yuh-Shan Ho. “Use of citation per publication as an indicator to evaluate contingent valuation research.” *CiteSeerX*, 12 April 2007, https://citeseerx.ist.psu.edu/viewdoc/download?doi=10.1.1.1040.7326&rep=rep1&type=pdf. Accessed 3 October 2022.

Little, Kendall. “What Is Blockchain and How Does It Work? | NextAdvisor with TIME.” *TIME*, 3 May 2022, https://time.com/nextadvisor/investing/cryptocurrency/what-is-blockchain/. Accessed 7 October 2022.

Marino, Lloyd. “Blockchain — The End of the Middleman | by Lloyd Marino.” *Medium*, 12 June 2016, https://medium.com/@LloydMarino/blockchain-the-end-of-the-middleman-37d97a67d7f. Accessed 6 October 2022.

Mercan, Suat, et al. “Improving transaction success rate in cryptocurrency payment channel networks.” *ScienceDirect*, 17 January 2019, https://www.sciencedirect.com/science/article/abs/pii/S0140366420320156. Accessed 10 November 2022.

Miau, Scott, and Jiann-Min Yang. “Bibliometrics-based evaluation of the Blockchain research trend: 2008 – March 2017.” *Technology Analysis and Strategic Management*, vol. 30, no. 9, 2018, pp. 1029-1045. *Taylor & Francis Online*, https://www.tandfonline.com/doi/full/10.1080/09537325.2018.1434138. Accessed 3 September 2022.

Milian, Eduardo Z., et al. “Fintechs: A literature review and research agenda.” *Electronic Commerce Research and Applications*, vol. 34, 2019. *SCOPUS*, www.scopus.com, Accessed 3 October 2022.

Moosavi, Javid, et al. “Blockchain in supply chain management: a review, bibliometric, and network analysis.” *Springer Link*, 27 February 2021, https://link.springer.com/article/10.1007/s11356-021-13094-3. Accessed 2 September 2022.

N-able. “SHA-256 Algorithm.” *N-able*, 12 September 2019, https://www.n-able.com/blog/sha-256-encryption. Accessed 5 October 2022.

Nasir, Adeel, et al. “What is Core and What Future Holds for Blockchain Technologies and Cryptocurrencies: A Bibliometric Analysis.” *IEEE Access*, vol. 9, 2021, pp. 989-1004. *IEEE Xplore*, https://ieeexplore.ieee.org/abstract/document/9305241. Accessed 6 October 2022.

Naved Khan, Mohammed, and Elena Kostadinova. “Determining mobile payment adoption: A systematic literature search and bibliometric analysis.” *Cogent Business & Management*, vol. 8, no. 1, 2021, pp. 1-21. *Taylor & Francis Group*, https://www.tandfonline.com/doi/pdf/10.1080/23311975.2021.1893245?needAccess=true. Accessed 3 September 2022.

Odoom, Justice, et al. “COVID-19 and future pandemics: A blockchain-based privacy-aware secure borderless travel solution from electronic health records.” *Wiley Online Library*, 22 July 2022, https://onlinelibrary.wiley.com/doi/full/10.1002/spe.3126. Accessed 7 October 2022.

Papadis, Nikolaos, and Leandros Tassiulas. “Blockchain-Based Payment Channel Networks: Challenges and Recent Advances.” *IEEE Access*, vol. 8, 2020, pp. 227596-227609, https://ieeexplore.ieee.org/abstract/document/9300150. Accessed 6 October 2022.

“Research collaborations bring big rewards: the world needs more.” *nature*, 16 June 2021, https://www.nature.com/articles/d41586-021-01581-z. Accessed 7 October 2022.

Rodeck, David. “What Is Blockchain? – Forbes Advisor.” *Forbes*, 28 April 2022, https://www.forbes.com/advisor/investing/cryptocurrency/what-is-blockchain/. Accessed 5 October 2022.

Sacred Heart University. “7. The Results - Organizing Academic Research Papers - Research Guides at Sacred

Heart University." *Sacred Heart University Library*, 7 September 2022, https://library.sacredheart.edu/c.php?g=29803&p=185931. Accessed 2 October 2022.

Salah, K., et al. "Blockchain for AI: Review and Open Research Challenges." *IEEE Access*, vol. 7, 2019, pp. 10127-10149. *SCOPUS*, www.scopus.com, doi:10.1109/ACCESS.2018.2890507.

Salah, Khaled. "Khaled Salah." *IEEE Xplore*, https://ieeexplore.ieee.org/author/37448520100. Accessed 7 October 2022.

Sarkar, Arijit. "Satoshi Nakamoto's Bitcoin white paper is now a 13-year-old teenager." *Cointelegraph*, 31 October 2021, https://cointelegraph.com/news/satoshi-nakamoto-s-bitcoin-white-paper-is-now-a-13-year-old-teenager. Accessed 5 October 2022.

"Scopus database & VOSViewer: Extracting data from Scopus for Bibliometric & scientometric analyses." *YouTube*, 13 April 2020, https://www.youtube.com/watch?v=azw1EEZrNI8. Accessed 4 October 2022.

"Self-citation and self-plagiarism, Research." *La Trobe University*, https://www.latrobe.edu.au/research/red/academic-integrity-for-researchers-and-ithenticate/self-citation-and-self-plagiarism. Accessed 3 October 2022.

Singh, Vivek K., et al. "The journal coverage of Web of Science, Scopus and Dimensions: A comparative analysis." *Springer Link*, 26 March 2021, https://link.springer.com/article/10.1007/s11192-021-03948-5. Accessed 2 October 2022.

Sonnenwald, Diane H. "Scientific Collaboration." *asis&t*, 24 October 2008, https://asistdl.onlinelibrary.wiley.com/doi/10.1002/aris.2007.1440410121. Accessed 4 October 2022.

Sterling, Jade. "Khalifa University Recognized as Most Published in Blockchain Oracle Research." *Khalifa University*, 28 November 2021, https://www.ku.ac.ae/khalifa-university-recognized-as-most-published-in-blockchain-oracle-research. Accessed 10 November 2022.

"Subject and Course Guides: Bibliometric Analysis and Visualization: Bibliometrics." *Subject and Course Guides*, 18 April 2022, https://researchguides.uic.edu/bibliometrics. Accessed 2 September 2022.

Sun, Ruo-Ting, et al. "Transformation of the Transaction Cost and the Agency Cost in an Organization and the Applicability of Blockchain—A Case Study of Peer-to-Peer Insurance." *Frontiers*, 27 April 2020, https://www.frontiersin.org/articles/10.3389/fbloc.2020.00024/full. Accessed 6 October 2022.

Tandon, Anushree, et al. "Blockchain applications in management: A bibliometric analysis and literature review." *Technological Forecasting and Social Change*, vol. 166, no. 166, 2021, pp. 1-19. *Elsevier*, https://www.sciencedirect.com/sdfe/reader/pii/S0040162521000810/pdf. Accessed 3 September 2022.

Tasatanattakool, Pinyaphat, and Chian Techapanupreeda. "Blockchain: Challenges and applications." *2018 International Conference on Information Networking (ICOIN)*, 2018, pp. 473-475. *IEEE Xplore*, https://ieeexplore.ieee.org/abstract/document/8343163?casa_token=ReG85ACgQZYAAAAA:bJ890ElVA8q5aEjjaBmh76q3wg3SIakpSYxuzoTnt4sS7M4Cp1_tN2uSaDh108qsJlyjh5AaDw. Accessed 6 October 2022.

Tchangalova, N., and J. Coalter. "Welcome - Systematic Review - Research Guides at University of Maryland Libraries." *Research Guides*, 28 August 2022, https://lib.guides.umd.edu/SR/welcome. Accessed 2 September 2022.

Thakor, A. V. "Fintech and Banking: What do we Know?" *Journal of Financial Intermediation*, vol. 41, 2020. *SCOPUS*, www.scopus.com, doi:10.1016/j.jfi.2019.100833.

"Using Smart Contracts for Automated Financial Features, Like Debt Repayment • Sila." *Sila Money*, 1 August 2022,

https://www.silamoney.com/ach/using-smart-contracts-for-automated-financial-features-like-debt-repayment. Accessed 6 October 2022.

van Eck, Jan, and Ludo Waltman. "[1109.2058] Text mining and visualization using VOSviewer." *arXiv*, 9 September 2011, https://arxiv.org/abs/1109.2058. Accessed 4 October 2022.

Viswanathan, Vibhuthi. "Implications of blockchain in data science." *ITProPortal*, 15 March 2022, https://www.itproportal.com/features/implications-of-blockchain-in-data-science/. Accessed 7 October 2022.

Wang, J., and H. Wang. *Monoxide: Scale Out Blockchain with Asynchronous Consensus Zones*, 2019. *SCOPUS*, www.scopus.com.

"What are 'publication citations'? : Dimensions." *Dimensions Support*, 19 September 2021, https://dimensions.freshdesk.com/support/solutions/articles/23000018845-what-are-publication-citations-. Accessed 3 October 2022.

"What Is Blockchain and How Does It Work?" *Synopsys*, https://www.synopsys.com/glossary/what-is-blockchain.html. Accessed 5 October 2022.

"What is Deterministic System? - Definition from Techopedia." *Techopedia*, https://www.techopedia.com/definition/602/deterministic-system. Accessed 5 October 2022.

"Why Blockchain Cannot Be Hacked..Or Can It?" *Bitrates.com*, https://www.bitrates.com/guides/blockchain/why-cannot-blockchain-be-hacked. Accessed 5 October 2022.

Zeng, Shuai, et al. "A Bibliometric Analysis of Blockchain Research." *IEEE Intelligent Vehicles Symposium*, 2018, pp. 102-107. *IEEE Xplore*, https://ieeexplore.ieee.org/abstract/document/8500606?casa_token=GtNAvi1GMKIAAAAA:TFblA_2wlZJEs3SbcneF1oE6mXHbdDrIcmBtuVwEsrctK6pvc3LhCuZWlZtFCayS2p8idByP1w. Accessed 6 October 2022.

Zicker, Fabio. "Co-authorship network analysis in health research: method and potential use - Health Research Policy and Systems." *Health Research Policy and Systems*, 30 April 2016, https://health-policy-systems.biomedcentral.com/articles/10.1186/s12961-016-0104-5#ref-CR1. Accessed 4 October 2022